\documentclass[aps,prb,twocolumn,superscriptaddress]{revtex4}
\usepackage{graphicx}
\usepackage{bm}
\usepackage{amsmath}

\begin{document} \title{Quantum vacuum properties of the intersubband cavity polariton
field}

\author{Cristiano Ciuti}
\email{ciuti@lpa.ens.fr} \affiliation{Laboratoire Pierre Aigrain,
Ecole Normale  Sup\'erieure, 24, rue Lhomond, 75005 Paris, France}

\author{G\'erald Bastard}
 \affiliation{Laboratoire Pierre Aigrain,
Ecole Normale  Sup\'erieure, 24, rue Lhomond, 75005 Paris, France}

\author{Iacopo Carusotto}
\affiliation{CRS BEC-INFM and Dipartimento di Fisica, Universit\`a
di Trento, I-38050 Povo, Italy}

\begin{abstract}
We present a quantum description of a planar microcavity photon
mode strongly coupled to a semiconductor intersubband transition
in presence of a two-dimensional electron gas. We show that, in
this kind of system, the vacuum Rabi frequency $\Omega_R$ can be a
significant fraction of the intersubband transition frequency
$\omega_{12}$. This regime of ultra-strong light-matter coupling
is enhanced for long wavelength transitions, because for a given
doping density, effective mass and number of quantum wells, the
ratio $\Omega_R/\omega_{12}$ increases as the square root of the
intersubband emission wavelength. We characterize the quantum
properties of the ground state (a two-mode squeezed vacuum), which
can be tuned {\it in-situ} by changing the value of $\Omega_R$,
e.g., through an electrostatic gate. We finally point out how the
tunability of the polariton quantum vacuum can be exploited to
generate correlated photon pairs out of the vacuum via quantum
electrodynamics phenomena reminiscent of the dynamical Casimir
effect.
\end{abstract}


\date{\today} \maketitle

In the last decade, the study of intersubband electronic
transitions\cite{ISB-Book} in semiconductor quantum wells has
enjoyed a considerable success, leading to remarkable
opto-electronic devices such as the quantum cascade
lasers\cite{Faist,Terahertz,Raffaele}. In contrast to the more
conventional interband transitions between conduction and valence
bands, the frequency of intersubband transitions is not
  determined by the energy gap
of the semiconductor material system used, but rather can be
chosen via the thickness of the quantum wells in the active
region, providing tunable sources emitting in the mid and far
infrared.

One of the most fascinating aspects of light-matter interaction is
the so-called strong light-matter coupling regime, which is
achieved when a cavity mode is resonant with an electronic
transition of frequency $\omega_{12}$, and the so-called vacuum
Rabi frequency $\Omega_R$ exceeds the cavity mode and electronic
transition linewidths. The strong coupling regime has been first
observed in the late '80s using atoms in metallic
cavities~\cite{atoms,Haroche}, and a few years later in
solid-state systems using excitonic transitions in quantum wells
embedded in semiconductor microcavities~\cite{Weisbuch}. In this
regime, the normal modes of the system consist of linear
superpositions of electronic and photonic excitations, which, in
the case of semiconductor materials, are the so-called {\em
polaritons}. In both these systems, the vacuum Rabi frequency
$\Omega_R$ does not exceed a very small fraction of the transition
frequency $\omega_{12}$.

Recently, Dini {\it et al.}\cite{Dini_PRL} have reported the first
demonstration of strong coupling regime between a cavity photon
mode and a mid-infrared intersubband transition, in agreement with
earlier semiclassical theoretical predictions by Liu\cite{Liu}.
The dielectric Fabry-Perot structure realized by Dini {\it et
al.}\cite{Dini_PRL} consists of a modulation doped multiple
quantum well structure embedded in a microcavity, whose mirrors
work thanks to the principle of total internal reflection. The
strong coupling regime has been also observed in quantum well
infra-red detectors\cite{Dupont}. As we will show in detail, an
important advantage of using intersubband transitions is the
possibility of exploring a regime where the normal-mode polariton
splitting is a significant fraction of the intersubband transition
(in the pioneering experiments by Dini {\it et
al.}\cite{Dini_PRL}, $2 \hbar \Omega_{R} = 14$ meV compared to
$\hbar \omega_{12} = 140$ meV). Furthermore, recent experiments
have also demonstrated the possibility of a dramatic tuning of the
strong light-matter coupling through application of a gate voltage
\cite{Tredicucci} which is able to deplete the density of the
two-dimensional electron gas.

Although the quest for quantum optical squeezing effects in the
emission from atoms strongly coupled to a cavity mode has been an
active field of research~\cite{CQED}, all systems realized up to
now show a vacuum Rabi frequency $\Omega_R$ much smaller than the
frequency of the optical transition. In this parameter regime, the
relative importance of the anti-resonant terms in the light-matter
coupling is small and, as far as no strong driving field is
present, they can be safely neglected under the so-called
rotating-wave approximation. In the presence of a strong driving
field, however, anti-resonant terms are known to play a
significant role, giving, e.g., the so-called Bloch-Siegert shift
in magnetic resonance experiments~\cite{BlochSiegert}, or
determining the quantum statistical properties of the emission
from dressed-state lasers~\cite{Lewenstein}.

A few theoretical studies have pointed out the intrinsic
non-classical properties of exciton-polaritons in solid-state
systems~\cite{Artoni,Schwendimann_bulk,Hradil}, but the small
value of the ratio $\Omega_R/\omega_{exc}$, typically less than
$0.01$, has so far prevented the observation of quantum effects
due to the anti-resonant terms of the light-matter coupling. All
the squeezing experiments that have been performed so far in fact
required the presence of a strong coherent optical pump beam in
order to inject polaritons and take advantage of nonlinear
polariton parametric
processes~\cite{CiutiPL,Schwendimann03,Karrsqueezing,KarrTwin,Ciuti04}.

In this paper, we show that in the case of {\em intersubband}
cavity polaritons, it is instead possible to achieve an
unprecedented {\em ultra-strong coupling} regime, in which the
vacuum Rabi frequency $\Omega_R$ is a large fraction of the
intersubband transition frequency $\omega_{12}$. To this purpose,
transitions in the far infrared are most favorable, because the
ratio $\Omega_R/\omega_{12}$ scales as the square root of the
intersubband emission wavelength. Within a second quantization
formalism, we characterize the polaritonic normal modes of the
system in the weak excitation limit, in which the density of
intersubband excitations is much smaller than the density of the
two-dimensional electron gas in each quantum well (in this very
dilute limit, the intersubband excitations behave as bosons). We
point out the non-classical properties of the ground state, which
consists of a two-mode squeezed vacuum. As its properties can be
modulated by applying an external electrostatic bias, we suggest
the possibility of observing quantum electrodynamics effects, such
as the generation of correlated photon pairs from the initial
vacuum state. Such an effect closely reminds the so-called
dynamical Casimir effect~\cite{CasDyn,friction,Lambrecht_review},
 whose observation is still an open challenge and is
actually the subject of intense effort. Many theoretical works
have in fact predicted the generation of photons in an optical
cavity when its properties, e.g. the length or the dielectric
permittivity of the cavity spacer material, are modulated in a
rapid, non-adiabatic way~\cite{Reynaud_PRL,DynCasRefr,Braggio}.

The present paper is organized as follows. In Sec.
\ref{sec:System} we describe the system under examination and in
Sec. \ref{sec:Hopfield} we introduce its Hamiltonian. The scaling
of the coupling intensity with the material parameters is
discussed in Sec. \ref{sec:scaling}, while Sec.
\ref{sec:Polaritons} is devoted to the diagonalization of the
Hamiltonian and the discussion of the polaritonic normal modes of
the system in the different regimes. The quantum ground state is
characterized in Sec. \ref{sec:ground} and its quantum properties
are pointed out. Two possible schemes for the generation of photon
pairs from the initial vacuum by modulating the properties of the
ground state are suggested in Sec. \ref{sec:casimir}. Conclusions
are finally drawn in Sec. \ref{sec:final}.

\section{Description of the system}
\label{sec:System}

\begin{figure}[t]
\begin{center}
\includegraphics[width=8.0cm]{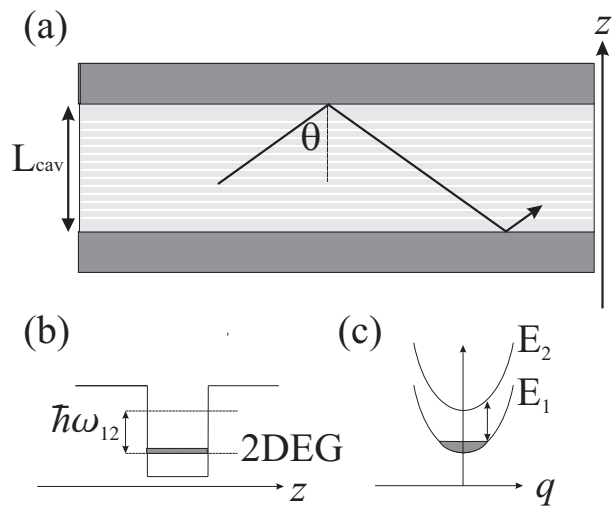}
\caption{ \label{Sketch_cavity} (a) Sketch of the considered
planar cavity geometry, whose growth direction is called $z$. The
cavity spacer of thickness $L_{\rm cav}$ embeds a sequence of
$n_{QW}$ identical quantum wells. The energy of the cavity mode
depends on the cavity photon propagation angle $\theta$. (b) Each
quantum well contains a two-dimensional electron gas in the lowest
subband (obtained through doping or electrical injection). The
transition energy between the first two subbands is $\hbar
\omega_{12}$. Only the TM-polarized photon mode is coupled to the
intersubband transition and a finite angle $\theta$ is mandatory
to have a finite dipole coupling. (c) Sketch of the energy
dispersion $E_1(q)$ and $E_2(q) = E_1(q) + \hbar \omega_{12}$ of
the first two subbands as a function of the in-plane wavevector
$q$. The dispersion of the inter-subband transition is negligible
as compared to the one of
  the cavity mode. For a typical value of the
cavity photon in-plane wavevector ${\bm k}$, one has in fact
$E_{2}(|{\bm k} + {\bm q}|) - E_{1}(q) \simeq \hbar \omega_{12}.$}
\end{center}
\end{figure}

In the following, we will consider a planar Fabry-Perot resonator
embedding a sequence of $n_{QW}$ identical quantum wells (see the
sketch in Fig. \ref{Sketch_cavity}). Each quantum well is assumed
to be doped with a two-dimensional density $N_{2DEG}$ of
electrons, which, at low temperatures, populate the first quantum
well subband. Due to the presence of the two-dimensional electron
gas, it is possible to have transitions from the first to the
second subband of the quantum well. We will call $\hbar
\omega_{12}$ the considered intersubband transition energy. If we
denote with $z$ the growth direction of the multiple quantum well
structure, then the dipole moment of the transition is aligned
along $z$, i.e., ${\bm d}_{12} = d_{12} \hat{\bm z}$. This
property imposes the well known polarization selection rule for
intersubband transitions in quantum wells, i.e., the electric
field must have a component along the growth direction. We point
out that in the case of a perfect planar structure, the in-plane
wave-vector is a conserved quantity, unlike the wave-vector
component along the $z$ direction. Therefore, all wave-vectors
${\bm k}$ will be meant as in-plane wave-vectors, unless
differently stated.

In the following, we will consider the fundamental cavity photon
mode, whose frequency dispersion is given by $ \omega_{{\rm
cav},k} = \frac{c}{\sqrt{\epsilon_{\infty}}} \sqrt{ k_z^2 + k^2}$,
where $\epsilon_{\infty}$ is the dielectric constant of the cavity
spacer and $k_z$ is the quantized photon wavevector along the
growth direction, which depends on the boundary conditions imposed
by the specific mirror structures. In the simplest case of
metallic mirrors, $k_z = \frac{\pi}{L_{\rm cav}}$, with $L_{\rm
cav}$ the cavity thickness.

\section{Second quantization Hamiltonian}

\label{sec:Hopfield} In this Section, we introduce the system
Hamiltonian in a second quantization formalism. In the following,
we will call $a^{\dagger}_{\bm k}$ the creation operator of the
fundamental cavity photon mode with in-plane wave-vector ${\bm
k}$. Note that, in order to simplify the notation, we will omit
the polarization index of the photon mode, which is meant to be
Transverse Magnetic (TM)-polarized (also known as p-polarization).
This photon polarization is necessary to have a finite value of
the electric field component along the growth direction $z$ of the
multiple quantum well structure, direction along which the
transition dipole of the intersubband transition is aligned.
$b^{\dagger}_{\bm k}$ will be instead the creation operator of the
bright intersubband excitation mode of the doped multiple quantum
well structure. In the simplified case of $n_{QW}$ identical
quantum wells that are identically coupled to the cavity photon
mode, the only bright intersubband excitation is the totally
symmetric one, with an oscillator strength $n_{QW}$ times larger
than the one of a single quantum well. The $n_{QW}-1$ orthogonal
excitations are instead dark and will be neglected in the
following. The creation operator corresponding to the bright
intersubband transition can be written as
\begin{equation}
\label{bdagger} b^{\dagger}_{\bm k} = \frac{1}{\sqrt{n_{QW}
N_{2DEG} S}} \sum_{j = 1}^{n_{QW}} \sum_{|\bm q| < k_F} c^{(j)
\dagger}_{2,{\bm q} +{\bm k}} c^{(j)}_{1,{\bm q}}~^,
\end{equation}
where $N_{2DEG}$ is the density of the two-dimensional electron
gas in each quantum well and $S$ is the sample area. The fermionic
operator $c^{(j)}_{1,{\bm q}}$ annihilates an electron belonging
to the first subband and $j$-th quantum well, while $c^{(j)
\dagger}_{2,{\bm q} +{\bm k}}$ creates an electron in the second
subband of the same well. $k_F$ is the Fermi wavevector of the
two-dimensional electron gas, whose electronic ground state at low
temperature is
\begin{equation}
|F\rangle = \prod_{j = 1}^{n_{QW}} \prod_{|\bm q| < k_F} c^{(j)
\dagger}_{1,{\bm q}} |0\rangle_{cond}~,
\end{equation}
where $|0\rangle_{cond}$ is the empty conduction band state.

In the following, we will consider the situation of a weakly
excited intersubband transition, i.e.,
\begin{equation}
\frac{1}{S}\,\sum_{\bm k}\langle b^{\dagger}_{\bm k} b_{\bm k}
\rangle \ll N_{2DEG}~.
\end{equation}
In this dilute limit, the intersubband excitation field is
approximately bosonic, namely
\begin{equation}
[b_{\bm k},b^{\dagger} _{\bm k'}] \simeq \delta_{{\bm k},{\bm
k'}}~.
\end{equation}

Starting from the coupled light-matter Hamiltonian of the
semiconductor and retaining only the considered cavity photon mode
for the electromagnetic field and the considered intersubband
transition for the electronic polarization field, one finds a
standard Hopfield-like Hamiltonian\cite{Hopfield}
\begin{equation}
H=H_0+H_{res}+H_{anti} \label{eq:Hamilt}
\end{equation}
which consists of three qualitatively different contributions,
namely
\begin{equation}
H_{0} =  \sum_{\bm{k}}  \hbar \omega_{{\rm cav},k} \left (
a^{\dagger}_{\bm{k}} a_{\bm{k}} + \frac{1}{2} \right ) + \sum_{\bm
k} \hbar \omega_{12}~b^{\dagger}_{\bm{k}} b_{\bm{k}}   ~,~
\label{H_0}
\end{equation}
\begin{multline}
H_{res} = \hbar \sum_{\bm{k}} \left \{ i \Omega_{R,k}  \left (
a^{\dagger}_{\bm{k}} b_{\bm{k}} - a_{\bm{k}}
b^{\dagger}_{\bm{k}}\right ) \right. \\
\left. + D_k \left ( a^{\dagger}_{\bm{k}} a_{\bm{k}} + a_{\bm{k}}
a^{\dagger}_{\bm{ k}} \right ) \right
 \}~,~ \label{H_res}
\end{multline}
\begin{multline}
H_{anti} = \hbar \sum_{\bm{k}} \left \{ i \Omega_{R,k}  \left (
a_{\bm{k}} b_{\bm{-k}} - a^{\dagger}_{\bm{k}} b^{\dagger}_{-
\bm{k}} \right ) \right. \\ \left.
 +  D_k \left ( a_{\bm{k}} a_{\bm{-k}}
+ a^{\dagger}_{\bm{k}} a^{\dagger}_{\bm{- k}} \right ) \right \}~.
\label{H_anti}
\end{multline}
$H_{0}$ in Eq. (\ref{H_0}) describes the energy of the bare cavity
photon and intersubband polarization fields, which depend on the
numbers $a^{\dagger}_{\bm{k}} a_{\bm{k}}$, $b^{\dagger}_{\bm{k}}
b_{\bm{k}}$ of cavity photons and intersubband excitations,
respectively.

$H_{res}$ in Eq. (\ref{H_res}) is the resonant part of the
light-matter interaction, depending on the vacuum Rabi energy
$\hbar \Omega_{R,k}$ and on the related coupling constant $D_k$.
The terms proportional to $\Omega_{R,k}$ describe the creation
(annihilation) of one photon and the annihilation (creation) of an
intersubband excitation with the same in-plane wavevector. In
contrast, the term proportional to $D_k$ contains only photon
operators, because it originates from the squared electromagnetic
vector potential part of the light-matter interaction.  Note that
this term in $H_{res}$ depends on the photon number operator
$a^{\dagger}_{\bm{k}} a_{\bm{k}}$ as the bare cavity photon term
in Eq. (\ref{H_0}). Hence, it gives a mere blueshift ($D_k > 0$)
of the bare cavity photon energy $\hbar \omega_{{\rm cav},k}$.

Finally, $H_{anti}$ in Eq. (\ref{H_anti}) contains the usually
neglected anti-resonant terms, which correspond to the
simultaneous destruction or creation of two excitations with
opposite in-plane wavevectors.  The terms proportional to
$\Omega_{R,k}$ describe the creation (or destruction) of a cavity
photon and an
  intersubband excitation, while the terms proportional to $D_k$
  describe the corresponding process involving a pair of cavity photons.

Before continuing our treatment, we wish to point out that the
considered Hamiltonian in Eq.(\ref{eq:Hamilt}) contains only the
energy associated to the fundamental cavity mode (including the
zero-point energy $\sum_{\bm k} \frac{1}{2} ~ \hbar \omega_{{\rm
cav},k}$), the energy associated to the creation of intersubband
excitations and the full light-matter interaction between the
considered modes. The energy terms associated to the other photon
modes, the electronic energy of the filled electronic bands as
well as the electrostatic energy associated to an applied bias
have been here omitted for simplicity, as they do not take part in
the dynamics discussed in
  the following of the paper.

\section{Scaling of the interaction}
\label{sec:scaling}

The specific values of the coupling constants $\Omega_{R,k}$ and
$D_k$ depend on the microscopic parameters of the intersubband
microcavity system.

The so-called vacuum Rabi energy $\hbar \Omega_{R,k}$ is the Rabi
energy obtained with the electric field corresponding to one
photon \cite{atoms,Cohen}. For the system under consideration
\cite{Dini_PRL,Liu}, the polariton coupling frequency for the
TM-polarized mode\cite{Andreani} reads
\begin{equation}
\Omega_{R,k} = \left ( \frac{2 \pi e^2}{\epsilon_{\infty}m_0
L^{\rm eff}_{\rm cav}} ~N_{2DEG}~n^{\rm eff}_{QW}f_{12} \sin^2
\theta \right  )^{1/2}~, \label{Rabi}
\end{equation}
where $\epsilon_{\infty}$ is the dielectric constant of the
cavity, $L^{\rm eff}_{\rm cav}$ the effective thickness of the
cavity photon mode (which depends non-trivially on the boundary
conditions imposed by the specific mirror structures), and $n^{\rm
eff}_{QW}$ the effective number of quantum wells ($n^{\rm
eff}_{QW} = n_{QW}$ in the case of quantum wells which are
identically coupled to the cavity photon field and which are
located at the antinodes of the cavity mode electric field). The
oscillator strength of the considered intersubband transition
reads
\begin{equation}
f_{12} = 2 m_0 \omega_{12} d_{12}^2/\hbar~,
\end{equation}
where $m_0$ is the free electron mass and $d_{12}$ is the
  electric dipole moment of the transition.
Under the approximation of a parabolic energy dispersion of the
  quantum well subbands, the oscillator strengths of the different
  intersubband transitions satisfy the $f$-sum
rule\cite{Sirtori94}
\begin{equation}
\sum_{j} f_{1j} = m_0/m^*,
\end{equation}
where $m^*$ is the effective electron mass of the conduction band.
In particular, for our case of a deep rectangular well, the sum
rule is almost saturated by the first intersubband transition
$f_{12} \simeq m_0/m^*$. Finally, $\theta$ is the propagation
angle {\it inside} the
  cavity (which is different from the propagation angle in the
  substrate), and is related to the in-plane wavevector $k$ by $k/k_z
  = \sin{\theta}/\cos{\theta}$.
\begin{figure}[t!]
\begin{center}
\includegraphics[width=8.0cm]{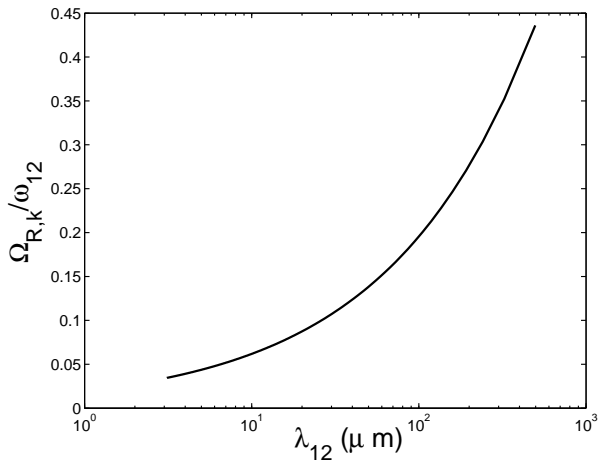}
\caption{\label{scaling_coupling} Coupling ratio
$\Omega_{R,k}/\omega_{12}$ as a function of the intersubband
emission wavelength $\lambda_{12}$ ($\mu$m). Parameters: $f_{12}=
12.9$ (GaAs quantum well), cavity spacer refraction index
$\sqrt{\epsilon_{\infty}} = 3$, $n^{\rm eff}_{QW} = 50$, $N_{2DEG}
= 5\times 10^{11}~{\rm cm}^{-2}$ and $\theta = 60^{\circ}$. The
Fabry-Perot resonator is a $\lambda/2$-microcavity. Results
obtained from the analytical expressions in Eqs. (\ref{ratio1})
and (\ref{ratio2}).}
\end{center}
\end{figure}

As we will see in the next section,  the relevant parameter
  quantifying the importance of the quantum effects
considered in this paper is the dimensionless ratio
$\Omega_{R,k_{\rm res}}/\omega_{12}$, where $k_{\rm res}$ is the
resonance in-plane wavevector such as $\hbar \omega_{{\rm
cav},k_{\rm res}} = \hbar \omega_{12}$. In the system studied by
Dini {\it et al.}\cite{Dini_PRL}, this ratio is already
significant, namely $\Omega_{R,k}/\omega_{12} = 0.05$. Here, we
show that the ratio $\Omega_{R,k}/\omega_{12}$ can be largely
increased designing structures in the far infra-red, by increasing
the number of quantum wells and by choosing semiconductors with
smaller effective mass.

Let be $\theta_{\rm res}$ the cavity propagation angle
corresponding to $k_{\rm res}$. From the relation
\begin{equation}
\label{k_res} k_{\rm res} = \frac{\omega_{12}}{c}
\sqrt{\epsilon_{\infty}} \sin \theta_{\rm res}~,
\end{equation}
we get that for metallic mirrors
\begin{equation}
L_{\rm cav}= \frac{\lambda_{12}}{2 \sqrt{\epsilon_{\infty}}
{\cos{\theta_{\rm res}}}}  ~,
\end{equation}
where $2\pi/{\lambda_{12}} = \omega_{12}/c$. Under these
conditions, the light-matter coupling ratio at the resonance angle
is
\begin{equation}
\frac{\Omega_{R,k_{\rm res}}}{\omega_{12}} =  \eta
\sqrt{\lambda_{12}}~, \label{ratio1}
\end{equation}
with
\begin{equation}
\eta = \sqrt { \frac{e^2~f_{12} \sin^2 \theta_{\rm res} \cos
\theta_{\rm res}~N_{2DEG}n_{QW}}{\pi m_{0} c^2
\sqrt{\epsilon_{\infty}}} } \label{ratio2}~.
\end{equation}
Note that the prefactor given in Eq.(\ref{ratio2}) has a weak
dependence on $\lambda_{12}$. In fact, in the limit case of a
rectangular quantum well with high potential barriers, $f_{12} =
0.96~m_0/m^{\star}$ and does not depend at all on $\lambda_{12}$.
More refined calculations \cite{Sirtori94} including the
non-parabolicity of the semiconductor band and the finite depth of
the potential well show that $f_{12}$ has a moderate dependence on
the emission wavelength $\lambda_{12}$ (it actually increases with
$\lambda_{12}$). Hence, the normalized vacuum Rabi frequency
$\Omega_{R,k_{\rm res}}/\omega_{12}$ increases at least as
$\sqrt{\lambda_{12}}$. The predictions of Eqs. (\ref{ratio1}) and
(\ref{ratio2}) are reported in Fig. \ref{scaling_coupling} for a
system of $50$ GaAs quantum wells and a doping density $N_{2DEG} =
5\times 10^{11}~{\rm cm}^{-2}$. For an intersubband emission
wavelength of $100 \mu{\rm m}$, the ratio $\Omega_R/\omega_{12}$
can be as high as $0.2$. The values in Fig. \ref{scaling_coupling}
can be significantly increased using semiconductors with smaller
effective mass, such as
InGaAs/AlInAs-on-InP~\cite{Sirtori_private}.

To complete our description, we need to provide the explicit
expression for the coefficient $D_k$, which quantifies the effect
of the squared electromagnetic vector potential in the
light-matter interaction. Generalizing Hopfield's
procedure~\cite{Hopfield} to the case of
  intersubband transitions, we find that all the intersubband
  transitions give a contribution to $D_k$, namely
\begin{equation}
D_{k} = \frac{\sum_{j} f_{1j}}{f_{12}}
\frac{\Omega_{R,k}^2}{\omega_{12}}~.
\end{equation}
However, as the oscillator strength of a deep rectangular well is
concentrated in the lowest transition at $\omega_{12}$, the effect
of the higher transitions is a minor correction, namely
\begin{equation}
D_{k} \simeq 1.04 \frac{\Omega_{R,k}^2}{\omega_{12}}\approx
\frac{\Omega_{R,k}^2}{\omega_{12}}.
\end{equation}
Note that for a quantum well with a parabolic confinement
potential $V(z) = (1/2) m^* \omega_{12}^2 z^2$, the expression
$D_{k} = \Omega_{R,k}^2/\omega_{12}$ would be exact, since in this
case all the intersubband oscillator strength is exactly
concentrated in the lowest transition $\omega_{12}$.

\section{Intersubband polaritons}
\label{sec:Polaritons}

\begin{figure}[t]
\begin{center}
\includegraphics[width=8.0cm]{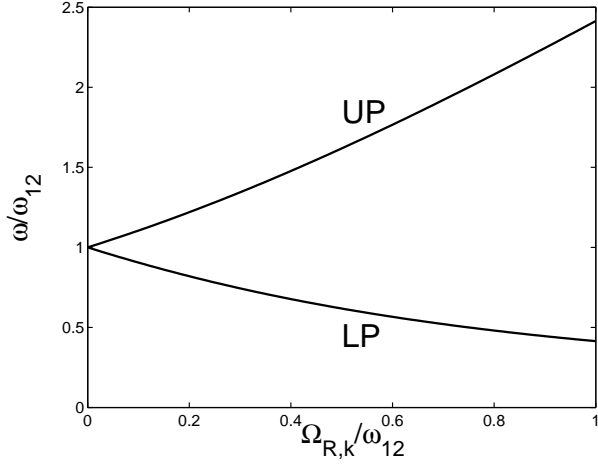}
\caption{ \label{Energies} Normalized polariton frequencies
$\omega_{LP,k}/\omega_{12}$ and $\omega_{UP,k}/\omega_{12}$ as a
function of $\Omega_{R,k}/\omega_{12}$ for $D_k =
\Omega_{R,k}^2/\omega_{12}$. The calculation has been performed
with $\omega_{{\rm cav},k} = \omega_{12}$. Note that for a given
microcavity system, $\Omega_{R,k}/\omega_{12}$ can be tuned {\it
in-situ} by an electrostatic bias, which is able to change the
density of the two-dimensional electron gas.}
\end{center}
\end{figure}

As all the terms in the Hamiltonian $H = H_{0} + H_{res} +
  H_{anti}$ are bilinear in the field operators, $H$
can be exactly diagonalized through a Bogoliubov transformation.
Following the pioneering work by Hopfield\cite{Hopfield}, we
introduce the Lower Polariton (LP) and Upper Polariton (UP)
annihilation operators
\begin{equation}
\label{p_Hopfield} p_{j, {\bm k}} = w_{j, k}~ a_{\bm k} +x_{j, k}~
b_{\bm k} + y_{j, k}~ a^{\dagger}_{-{\bm k}} + z_{j, {k}}~
b^{\dagger}_{-\bm k}~,
\end{equation}
where $j \in \{LP, UP\}$. The Hamiltonian of the system can be
cast in the diagonal form
\begin{equation}
H =  E_G + \sum_{{j \in \{LP, UP\}}}  \sum_{\bm{k}} \hbar
\omega_{j,k}~ p^{\dagger}_{j,\bm{k}} p_{j,\bm{k}}  ~,~
\label{diagonal}
\end{equation}
where the constant term $E_G$ will be given explicitly later. The
Hamiltonian form in Eq. (\ref{diagonal}) is  obtained, provided
that the vectors
\begin{equation}
\vec{v}_{j,k} = (w_{j, k},x_{j, k},y_{j, k},z_{j, k})^T
\end{equation}
satisfy the eigenvalues equation
\begin{equation}
M_k \vec{v}_{j,k} = \omega_{j,k} \vec{v}_{j,k}
\end{equation}
with $\omega_{j,k} > 0$. The Bose commutation rule
\begin{equation}
[p_{j,{\bm k}},p^{\dagger}_{j',{\bm k'}}] = \delta_{j,j'}
\delta_{{\bm k}, {\bm k'}}
\end{equation}
 imposes the normalization condition
\begin{equation}
w_{j,k}^* w_{j',k}+ x_{j,k}^* x_{j',k} - y_{j,k}^* y_{j',k} -
z_{j,k}^* z_{j',k}= \delta_{j,j'}~.
\end{equation}

The Hopfield-like matrix for our system reads
\begin{equation}
\label{Hopfieldmatrix} M_k = \left (
\begin{array}{cccc}
\omega_{{\rm cav},k} + 2 D_k  & -i \Omega_{R,k} & -2D_k & -i \Omega_{R,k} \\
i\Omega_{R,k} & \omega_{12} & - i\Omega_{R,k} & 0 \\
2D_k & -i\Omega_{R,k} & -\omega_{{\rm cav},k} - 2 D_k  & -i \Omega_{R,k} \\
-i\Omega_{R,k} & 0 & i\Omega_{R,k} &  -\omega_{12} \\
\end{array}
\right )~.
\end{equation}
The four eigenvalues of $M_k$ are $\{\pm \omega_{LP,k},\pm
\omega_{UP,k}\}$. Under the approximation $D_k =
\Omega_{R,k}^2/\omega_{12}$ (i.e., all the oscillator strength
concentrated on the $\omega_{12}$ transition), $\det{M_k} =
(\omega_{{\rm cav},k}~\omega_{12})^2$, giving the simple relation
\begin{equation}
\omega_{LP,k}~ \omega_{UP,k} = \omega_{12}~\omega_{{\rm cav},k}~,
\label{geometric}
\end{equation}
i.e., the geometric mean of the energies of the two polariton
branches is equal to the geometric mean of the bare intersubband
and cavity mode energies. The dependence of the exact polariton
eigenvalues as a function of $\Omega_{R,k}/\omega_{12}$ is
reported in Fig. \ref{Energies}, for the resonant case
$\omega_{{\rm cav},k} = \omega_{12}$.
\begin{figure}[t!]
\begin{center}
\includegraphics[width=8cm]{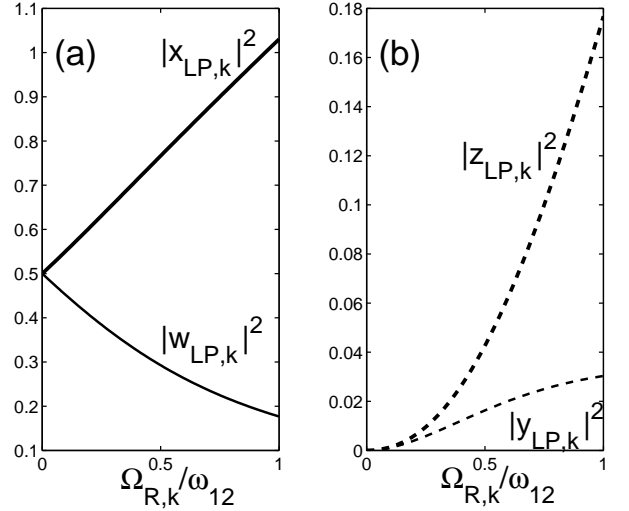}
\caption{ \label{LP_mixing_fractions} Mixing fractions for the
Lower Polariton (LP) mode as a function of
$\Omega_{R,k}/\omega_{12}$ (see Eq. (\ref{p_Hopfield}) in the
text). The calculation has been performed for the resonant case
$\omega_{{\rm cav},k} = \omega_{12}$ , as in the previous figure.
Panel (a): $|w_{LP,k}|^2$ (thin solid line), $|x_{LP,k}|^2$ (thick
solid line). Note that for $\Omega_{R,k}/\omega_{12} \ll 1$,
$|w_{LP,k}|^2 \simeq |x_{LP,k}|^2 \simeq 1/2$. Panel (b):
$|y_{LP,k}|^2$ (thin dashed line), $|z_{LP,k}|^2$ (thick dashed
line). For $\Omega_{R,k}/\omega_{12} \ll 1$, $|y_{LP,k}|^2 \simeq
|z_{LP,k}|^2 \simeq 0$. The Upper Polariton (UP) fractions (not
shown) are simply $|w_{UP,k}|^2 = |x_{LP,k}|^2$, $|x_{UP,k}|^2 =
|w_{LP,k}|^2$, $|y_{UP,k}|^2 = |z_{LP,k}|^2$, $|z_{UP,k}|^2 =
|y_{LP,k}|^2$.}
\end{center}
\end{figure}

\subsection{Ordinary properties in the limit $\Omega_{R,k}/\omega_{12}
\ll 1$}

In the standard case $\Omega_{R,k}/\omega_{12} \ll 1$, the
polariton operator can be approximated as
\begin{equation}
p_{j, {\bm k}} \simeq w_{j, k}~ a_{\bm k} +x_{j,  k}~ b_{\bm k} ~,
\label{p_operators}
\end{equation}
with $|w_{j,k}|^2 + |x_{j,k}|^2 \simeq 1$. This means that the
annihilation operator for a polariton mode with in-plane
wave-vector ${\bm k}$ is given by a linear superposition of the
photon and intersubband excitation annihilation operators with the
same in-plane wavevector, while mixing with the creation operators
(represented by the coefficients $y_{j,k}$ and $z_{j,k}$) is
instead negligible [see Fig. \ref{LP_mixing_fractions}]. In this
limit, the geometric mean can be approximated by the arithmetic
mean and Eq. (\ref{geometric}) can be written in the more usual
form:
\begin{equation}
\omega_{LP,k} + \omega_{UP,k} \simeq \omega_{{\rm cav},k} +
\omega_{12}~.
\end{equation}
For the specific resonant wavevector $k_{\rm res}$ such that
  $\omega_{{\rm cav},k_{\rm res}} = \omega_{12}$, the polariton eigenvalues
are
\begin{equation}
\omega_{LP(UP),k_{\rm res}} \simeq \omega_{12} \mp
\Omega_{R,k_{\rm res}}~,
\end{equation}
and the mixing fractions are $|w_{LP,k_{\rm res}}|^2 \simeq
|x_{LP,k_{\rm res}}|^2 \simeq 1/2$.

\subsection{Ultra-strong coupling regime}
\label{strong}

When the ratio $\Omega_{R,k}/\omega_{12}$ is not negligible
compared to $1$, then the anomalous features due to the
anti-resonant terms of the light-matter coupling becomes truly
relevant.

In the resonant $\omega_{{\rm cav},k_{\rm res}}=\omega_{12}$ case
and under the approximation $D_k = \Omega_{R,k}^2/\omega_{12}$,
the polariton frequencies are given by
\begin{eqnarray}
\omega_{LP(UP),k_{\rm res}} = \sqrt{\omega_{12}^2+
(\Omega_{R,k_{\rm res}})^2} \mp \Omega_{R,k_{\rm res}},
\end{eqnarray}
which, as it is apparent in Fig. \ref{Energies}, corresponds to a
strongly asymmetric anti-crossing as a function of
$\Omega_{R,k_{\rm res}}/\omega_{12}$. This is due to the combined
effect of the blue-shift of the cavity mode frequency due to the
terms proportional to $D_k$ in Eq. (\ref{H_res}), and of the
anomalous coupling terms in Eq.(\ref{H_anti}).

These same effects contribute to the non-trivial evolution of the
  Hopfield coefficients shown in Fig. \ref{LP_mixing_fractions}.
The anomalous Hopfield fractions $|y_{LP,k}|^2$ and $|z_{LP,k}|^2$
  significantly increase because of the anomalous coupling, and
  eventually become of the same order as the normal ones
  $|x_{LP,k}|^2$ and $|w_{LP,k}|^2$.
Due to the normalization condition
\begin{equation}
|w_{j,k}|^2 + |x_{j,k}|^2 - |y_{j,k}|^2 - |z_{j,k}|^2 = 1~,
\end{equation}
this affects the ordinary fractions $|w_{LP,k}|^2$, $|x_{LP,k}|^2$
as well. Owing to the blue-shift of the cavity photon frequency
induced by the light-matter coupling, at the resonance wavevector
$k= k_{\rm res}$ the lower
  polariton becomes more matter-like (i.e., $ |x_{LP,k_{\rm res}}|^2 >
  |w_{LP,k_{\rm res}}|^2$ and $|z_{LP,k_{\rm res}}|^2 > |y_{LP,k_{\rm res}}|^2$),
  while the upper polariton more photon-like. In this resonant case, the UP Hopfield coefficients (not shown)
are simply related to the LP ones by: $|w_{UP,k_{\rm res}}|^2 =
|x_{LP,k_{\rm res}}|^2$, $|x_{UP,k_{\rm res}}|^2 = |w_{LP,k_{\rm
res}}|^2$, $|y_{UP,k_{\rm res}}|^2 = |z_{LP,k_{\rm res}}|^2$,
$|z_{UP,k_{\rm res}}|^2 = |y_{LP,k_{\rm res}}|^2$.

\section{The quantum ground state}
\label{sec:ground}

\subsection{The normal vacuum state $|0\rangle$ for $\Omega_{R,k} = 0$}
In the case $\Omega_R = 0$ (negligible light-matter interaction),
the quantum ground state $|G\rangle$ of the considered system is
the ordinary vacuum $|0\rangle$ for the cavity photon and
intersubband excitation fields. Such ordinary vacuum satisfies the
relation
\begin{equation}
a_{\bm k} |0\rangle = b_{\bm k} |0\rangle = 0~,
\end{equation}
which means a vanishing number of photons and intersubband
excitations:
\begin{equation}
 \langle 0 |a^{\dagger}_{\bm k} a_{\bm k} |0\rangle =
\langle 0 |b^{\dagger}_{\bm k} b_{\bm k} |0\rangle =  \langle 0
|a^{\dagger}_{\bm k} b_{\bm k} |0\rangle = 0
\end{equation}
and no anomalous correlations, i.e.,
\begin{equation}
\langle 0 |a_{\bm k} a_{\bm k'} |0\rangle = \langle 0 |b_{\bm k}
b_{\bm k'} |0\rangle = \langle 0 |a_{\bm k} b_{\bm k'} |0\rangle =
0~.
\end{equation}

\subsection{The squeezed vacuum state}
With a finite $\Omega_{R,k}$, the ground state of the system
$|G\rangle$ is no longer the ordinary vacuum $|0\rangle$ such
that:
\begin{equation}
a_{\bm k}|0\rangle=b_{\bm k}|0\rangle=0~,
\end{equation}
but rather the vacuum of polariton excitations:
\begin{equation}
p_{j,\bm k} |G\rangle = 0~. \label{squeezed}
\end{equation}
As the polariton annihilation operators are linear superpositions
of annihilation and {\em creation} operators for the photon and
the intersubband excitation modes, the ground state $|G\rangle$
is, in quantum optical terms, a squeezed
state~\cite{Slusher,Walls}. By inverting Eq.(\ref{p_Hopfield}),
one gets
\begin{widetext}
\begin{equation}
\left (
\begin{array}{c}
a_{\bm k} \\
b_{\bm k} \\
a^{\dagger}_{-\bm k} \\
b^{\dagger}_{-\bm k} \\
\end{array}
\right ) = \left (
\begin{array}{cccc}
w^*_{LP, k} & w^*_{UP, k} & - y_{LP, k} &  - y_{UP,
k} \\
x^*_{LP, k} & x^*_{UP, k} & - z_{LP, k} &  - z_{UP,
k}\\
-y^*_{LP, k} & -y^*_{UP, k} & w_{LP, k} &  w_{UP, k} \\
-z^*_{LP, k} & -z^*_{UP, k} & x_{LP, k} &  x_{UP, k}
\end{array}
\right )
 \left (
\begin{array}{c}
p_{LP,\bm k} \\
p_{UP,\bm k} \\
p^{\dagger}_{LP,-\bm k} \\
p^{\dagger}_{UP,-\bm k} \\
\end{array}
\right )~,
\end{equation}
\end{widetext}
from which, using Eq. (\ref{squeezed}) and the boson commutation
rules, we obtain that the ground state contains a finite number
(per mode) of
  cavity photons and intersubband excitations:
\begin{eqnarray}
\langle G |a^{\dagger}_{\bm k} a_{\bm k} |G\rangle &=&  |y_{LP,
k}|^2 + |y_{UP, k}|^2  \label{n_phot} \\
 \langle G
|b^{\dagger}_{\bm k} b_{\bm k} |G\rangle &=& |z_{LP, k}|^2 +
|z_{UP, k}|^2~\label{n_exc},
\end{eqnarray}
as well as some correlation between the photon and intersubband
fields:
\begin{equation}
\label{ab} \langle G |a^{\dagger}_{\bm k} b_{\bm k} |G\rangle =
y^{*}_{LP, k} z_{LP, k} + y^{*}_{UP, k} z_{UP, k}~.
\end{equation}
Moreover, significant anomalous correlation exist between opposite
momentum components of the fields:
\begin{eqnarray}
\label{aa} \langle G |a_{\bm k} a_{-\bm k} |G\rangle &=& -
w^*_{LP,k} y_{LP, k} - w^*_{UP,k} y_{UP, k} \\
\label{bb} \langle G |b_{\bm k} b_{-\bm k} |G\rangle &=& -
x^*_{LP,k} z_{LP, k} - x^*_{UP,k} z_{UP, k} \\
\label{ab_an} \langle G |b_{\bm k} a_{-\bm k} |G\rangle &=& -
x^*_{LP,k} y_{LP, k} - x^*_{UP,k} y_{UP, k}~.
\end{eqnarray}
Note that the finite photonic population which is present in the
ground state $|G\rangle$ of our system is composed of "virtual"
photons. In the absence of any perturbation or modulation of the
parameters of the quantum Hamiltonian, these virtual photons can
not escape from the cavity and therefore do not result in any
observable emitted radiation (indeed, energy would not be
conserved in such a process).

\begin{figure}[t]
\begin{center}
\includegraphics[width=8cm]{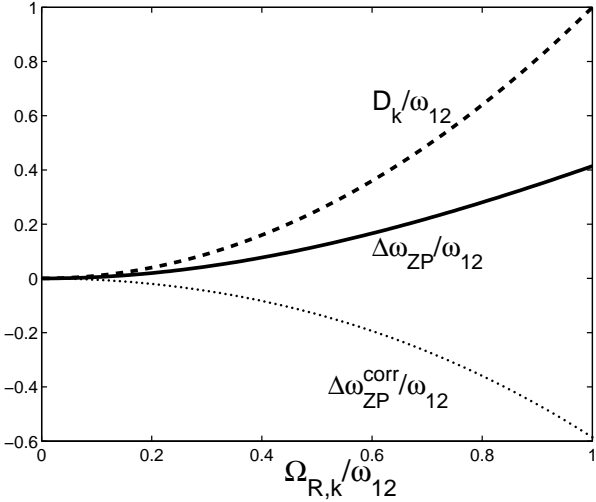}
\caption{Solid line: normalized differential zero-point energy
(per mode) $\Delta \omega_{ZP}(k)/\omega_{12}$ as a function of
$\Omega_{R,k}/\omega_{12}$. Dashed line: $D_k/\omega_{12}$. Dotted
line: normalized correlation contribution $\Delta
\omega_{ZP}^{corr}(k)/\omega_{12}$. The calculation has been
performed with $\omega_{{\rm cav},k} = \omega_{12}$. \label{ZPE}}
\end{center}
\end{figure}
As it has been shown in Fig.~\ref{LP_mixing_fractions}, the
Hopfield coefficients $x_{j,k},y_{j,k},w_{j,k},z_{j,k}$ as well as
the ground state $|G\rangle$ of the system strongly depend on the
vacuum Rabi energy $\Omega_{R,k}$, which in our case can be
dramatically modulated {\em in situ}, e.g. by changing the
electron density $N_{2DEG}$ via a time-dependent external
electrostatic bias. In particular, we shall discuss how this
remarkable tunability of the system can be used to "unbind" the
virtual photons by modulating the parameters of the system in a
time-dependent way, and generate some radiation which can be
actually detected outside the cavity. These issues will be the
subject of sec. \ref{sec:casimir}.

\subsection{The ground state energy}

Also the energy $E_G$ of the quantum ground state has a
significant
  dependence on the coupling $\Omega_{R,k}$.
Defining $E_0$ as the ground state energy of the uncoupled
  ($\Omega_{R,k}=0$) system, we have that:
\begin{equation}
E_G - E_0 = \sum_{\bm k} \big[\hbar D_k  - \sum_{j \in \{ LP,UP
\}} \hbar \omega_{j,k} (|y_{j, k}|^2 + |z_{j, k}|^2)\big ]~.
\label{eq:E_G}
\end{equation}
Note that this energy difference includes only the contribution of
the zero-point fluctuations of the intersubband polariton field
and does not take into account the other contributions coming,
e.g. from the change of the electrostatic energy of the system
(which is imposed by an applied bias), as already discussed at the
end of Sec. \ref{sec:Hopfield}. The (always positive) term $D_k$
in Eq.(\ref{eq:E_G}) is the zero-point energy change due to the
mere blue-shift of the bare cavity mode frequency and does not
correspond to any squeezing effect. The second term is instead due
to the mixing of creation and annihilation operators into the
polaritonic operators as described in Eq.(\ref{p_Hopfield}) and is
proportional to the number of virtual photons and intersubband
excitations present in the ground state $|G\rangle$ of the system
according to Eqs.(\ref{n_phot}-\ref{n_exc}). As it is usual for a
correlation contribution, it tends to lower the ground state
energy. It is interesting to study the differential "zero-point"
energy per mode $\hbar \Delta \omega_{ZP}(k)$, whose sum over all
the ${\bm k}$-modes gives the quantum ground state energy
difference $E_G-E_0$. The differential "zero-point" frequency
reads
\begin{equation}
\Delta \omega_{ZP}(k) = D_k + \Delta \omega_{ZP}^{corr}(k)~,
\end{equation}
where the (negative) correlation contribution reads
\begin{equation}
\Delta \omega_{ZP}^{corr}(k) =  - \sum_{j \in \{LP,UP \}}
\omega_{j,k}~(|y_{j, k}|^2 + |z_{j, k}|^2)~.
\end{equation}
 These quantities (normalized to $\omega_{12}$) are plotted in Fig. \ref{ZPE} as a function of
$\Omega_{R,k}/\omega_{12}$ for the resonant case $\omega_{{\rm
cav},k} = \omega_{12}$. Although it is the diagonal blueshift
which gives the dominant contribution to the ground state energy
shift, the negative contribution due to the correlation effects is
important, being as large as $- 0.13 \hbar \omega_{12}$ already
for $\Omega_{R,k}/\omega_{12} = 0.5$.

\section{Tuning the quantum vacuum: quantum radiation effects}
\label{sec:casimir}

\begin{figure}[t]
\begin{center}
\includegraphics[width=8cm]{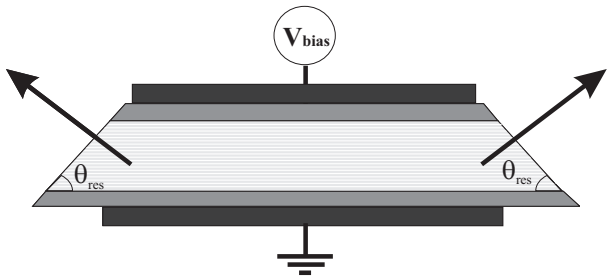}
\caption{Sketch of a possible set-up for the generation of
correlated photon pairs in the intersubband cavity system. The
vacuum Rabi frequency of the intersubband cavity system can be
modulated through an electric gate, which changes the density of
the two-dimensional electron gas or alternatively the dipole
moment of the intersubband transition. A modulation of the bias is
expected to induce the emission of correlated photon pairs with
opposite in-plane wavevectors. This kind of radiation can be
optimally guided out of the cavity through wedged lateral facets,
with inclination equal to the resonance angle $\theta_{\rm res}$.
\label{Twin_photon}}
\end{center}
\end{figure}

The possibility of tuning in a dramatic way the properties (energy
and squeezing) of the ground state of the system, as well as the
significant number of (virtual) excitations already present in the
ground state suggests that the present system could be a potential
laboratory to study Quantum Electro-Dynamics (QED) phenomena,
which are reminiscent of the dynamic Casimir
effects\cite{CasDyn,friction,Lambrecht_review}. In particular, we
shall discuss how a time-modulation of the ground-state properties
of the system can parametrically produce {\em real} excitations
above the ground state of our cavity, which then escape from the
cavity as photons and propagate in the external free-space.

In the typical arrangement for the observation of the dynamical
Casimir effect, one has to modulate in time the properties of an
optical cavity and, in particular, its resonance frequencies.
Several proposals have appeared in order to do this: in the
simplest ones, one has to periodically move the mirrors so as to
modify the boundary conditions of the
field~\cite{Lambrecht_review,Reynaud_PRL}. Other
proposals~\cite{DynCasRefr} deal with a time-dependance of the
refractive index of a dielectric medium placed inside the cavity.
A recent work proposes to vary the effective length of the cavity
by changing the reflectivity of a composite mirror~\cite{Braggio}.

The main peculiarity of our system as compared to previous
proposals is due to the possibility of modulating the properties
of the ground state in a much stronger way, due to the
ultra-strong and tunable light-matter coupling.

In the next subsection, we shall give a detailed analysis of a
simple {\em gedanken} experiment, where the vacuum Rabi frequency
is assumed to be switched off in an instantaneous way. This scheme
has the merit of allowing to grasp the essential physics of the
problem, providing quantitative estimates without the need of
embarking in complicate calculations.

A complete and quantitative calculation of the spectral shape and
intensity of the emitted radiation for the most relevant case of a
periodic modulation of $\Omega_{R,k}$ is beyond the scope of the
present paper, as it would require a careful analysis of the
coupling of the cavity system to the extra-cavity field as well as
of the other non-radiative loss mechanisms of the electronic
system\cite{Robson}. This is work actually in progress and here we
shall restrict ourselves to a very qualitative discussion of its
main features.

\subsection{Abrupt switch off of the vacuum Rabi energy}

Let us suppose that the considered intersubband cavity system is
in the ground state $|G\rangle$. As we have already discussed, the
squeezed vacuum $|G\rangle$ contains a finite number of cavity
photons and intersubband excitations because of the correlations
due to the anomalous coupling terms in Eq.(\ref{H_anti}).

If one switches off the vacuum Rabi frequency $\Omega_{R,k}$ of
the system in an abrupt, non-adiabatic way by suddenly depleting
the electron gas, the photon mode does not have the time to
respond to the perturbation and will remain in the same squeezed
vacuum state as before. As this state is now an excited state of
the Hamiltonian for $\Omega_{R,k}=0$, the system will relax
towards its ground state, which now corresponds to the standard
vacuum, by emitting the extra photons as propagating radiation.

One possible way to collect this {\em quantum vacuum radiation} is
through the set-up sketched in Fig. \ref{Twin_photon}, which
allows one to collect the photons which are emitted with internal
propagation angle $\theta$ around the resonance value
$\theta_{res}$. If one neglects the losses due to the background
absorption by the dielectric material forming the microcavity, an
estimate of the number of emitted photons can be obtained as
follows. The number of photon states (per unit area) in the 2D
momentum volume $d^2{\bm k}$ is simply $d^2{\bm k}/(2\pi)^2$.
Hence, the differential density of photons (per unit area) in the
2D momentum volume $d^2{\bm k}$ is
\begin{equation}
d \rho_{phot}  = \frac{d^2{\bm k}}{(2 \pi)^2} \langle G |
a^{\dagger}_{\bm k} a_{\bm k} |G\rangle~,
\end{equation}
where the photon number $\langle G | a^{\dagger}_{\bm k} a_{\bm k}
|G\rangle$ in the quantum ground state is given by Eq.
(\ref{n_phot}). Now, all the expectation values depend only on
$|{\bm k}|$ and hence we can rewrite the momentum volume as
$d^2{\bm k} = 2 \pi k dk$. Knowing that the in-plane wavevector
$k$ is given by the relationship $k = k_z \tan(\theta)$ and using
Eq. (\ref{k_res}), we find the final result
\begin{equation}
\label{pulse} \frac{d \rho_{phot}}{d \theta} (\theta_{res}) =
\frac{1}{2 \pi} \frac{\omega_{12}^2}{c^2} \epsilon_{\infty}
\tan(\theta_{res}) \langle G | a^{\dagger}_{\bm k} a_{\bm k}
|G\rangle~.
\end{equation}

To give a numerical application of Eq. (\ref{pulse}), let us
consider an intersubband cavity system with $\hbar \omega_{12} =
140$ meV, resonance angle $\theta_{\rm res} = 65^{\circ}$ and
$\hbar \Omega_{R,k_{\rm res}} = 7$ meV (these are approximately
the values in the sample measured by Dini {\it et
al}\cite{Dini_PRL}). For these parameters, Eq. (\ref{pulse}) gives
the differential photon density $d\rho_{phot}/d\theta \simeq 1
\times 10^5$ cm$^{-2}$ rad$^{-1}$.

Note that the emission corresponding to the ${\bm k}$-mode is
correlated to the emission corresponding to the mode with opposite
in-plane wavevector, as shown in Eq. (\ref{aa}). Indeed, the
"quantum vacuum radiation" here described consists in the emission
of correlated photon pairs\cite{Treps}.

\subsection{Periodic modulation of $\Omega_{R,k}$}

The requirement of a abrupt, non-adiabatic, switch-off of the
  Rabi coupling $\Omega_{R,k}$ imposes very stringent limits on the time-scale
$\tau_{sw}$  over which the electrostatic bias has to be applied.
In particular, we expect that in order to maximize the quantum
vacuum radiation generation, $\tau_{sw}$ can not be too much
longer that the oscillation
  period of the lower polaritonic mode.

It is then perhaps more accessible from an experimental point of
view to try to
  detect the vacuum radiation by periodically modulating the vacuum
  Rabi frequency at an
  angular frequency $\omega_{mod}$
\begin{equation}
\Omega_{R,k}(t) = \bar{\Omega}_{R,k} + \Delta{\Omega}_{R,k}
\sin(\omega_{mod} t).
\end{equation}
Note that in principle this kind of modulation can be obtained not
only through a gate-induced depletion of the two-dimensional
electron gas\cite{Tredicucci}, but also by modulating the dipole
moment of the intersubband transition or alternatively the
reflectivity of the mirrors.
 As all the relevant physical quantities in the present
problem (polariton energies, Hopfield coefficients, ground state
energy) depend in a nonlinear way on the vacuum Rabi frequency
$\hbar \Omega_{R,k}$, we expect that for large modulation
amplitudes high order harmonics of the fundamental modulation
frequency $\omega_{mod}$ will play a significant role in the
parametric process which is responsible for the vacuum radiation
generation~\cite{Lambrecht_review}. In particular, emission will
be enhanced if
\begin{equation}
\omega_{j,{\bm k}} + \omega_{j',-{\bm k}} = r~ \omega_{mod}~,
\end{equation}
with $r$ being a generic positive integer number, and
$j,j'=\{LP,UP\}$. This is the phase-matching condition for the
parametric generation of two polaritons with opposite momentum. As
usual, the narrower the polaritonic resonance, the stronger the
resonant enhancement.

As it is generally the case for parametric processes in a cavity,
the number of photons which are generated in the cavity and then
emitted as radiation is determined by a dynamical equilibrium
between the parametric processes generating them and the losses,
the radiative as well as the non-radiative ones\cite{Robson}. For
a complete and quantitative treatment of these issues, further
investigations are in progress.

\section{Conclusions}
\label{sec:final}

In conclusion, we have shown that in the intersubband cavity
polariton system, a new regime of ultra-strong coupling can be
  achieved, where
the vacuum Rabi frequency $\Omega_R$ is a large fraction of the
intersubband transition frequency $\omega_{12}$. This scenario
appears to be easier to achieve
  in the far infrared, since the ratio $\Omega_R/\omega_{12}$
scales as the square root of the intersubband transition
wave-length. In the ultra-strong coupling regime, the usually
neglected anti-resonant terms of the light-matter coupling start
playing an important role. In particular, the ground state of
system is no longer the ordinary vacuum of photons and electronic
excitations, but rather a
  two-mode squeezed vacuum, whose properties strongly depend on the
  ratio $\Omega_R/\omega_{12}$. As this quantity can be dramatically tuned by applying an
electrostatic bias, we have pointed out the possibility of
observing interesting  quantum electrodynamic effects reminiscent
of the dynamical Casimir effect, i.e. the generation of correlated
photon pairs out of the initial polariton vacuum state.  A
quantitative estimate of the number of emitted photons has been
given for the simplest case of an instantaneous switch-off of the
light-matter coupling, and the results look promising in view of
experimental observations. Work is actually in progress in the
direction of extending the analysis to the case of a periodic
modulation of $\Omega_{R,k}$, case in which one should be able to
enhance the emitted intensity via parametric resonance effects.
From the theoretical point of view, this study requires a complete
treatment of losses in order to describe the dynamical equilibrium
between the parametric process generating the quantum radiation
and the dissipation.

\acknowledgments It is our pleasure to thank Raffaele Colombelli
and Carlo Sirtori for many enthusiastic and stimulating meetings
about intersubband photonics. CC would like to thank Alessandro
Tredicucci and A. Anappara for showing data concerning the
gate-controlled vacuum Rabi energy prior to publication. We are
grateful to Paolo Schwendimann, Antonio Quattropani, Arnaud
Verger, Chiara Menotti, Alessio Recati, Mauro Antezza and Maurizio
Artoni for discussions and/or for a critical reading of this
manuscript. LPA-ENS is a "Unit\'{e} Mixte de Recherche Associ\'{e}
au CNRS (UMR 8551) et aux Universit\'{e}s Paris 6 et 7".

\end{document}